# Dynamics of Graphene/Al Interfaces using COMB3 Potentials

Difan Zhang[1,2], Alexandre F. Fonseca[3], Tao Liang[2], Simon R. Phillpot[1], Susan B. Sinnott[2,4,5]*

[1]Department of Materials Science and Engineering, University of Florida, Gainesville, FL, 32611, USA

[2]Department of Materials Science and Engineering, The Pennsylvania State University, University Park, PA, 16802, USA

[3]Applied Physics Department, State University of Campinas, Campinas, SP, 13083-970, Brazil

[4]Materials Research Institute, The Pennsylvania State University, University Park, PA, 16802, USA

[5]Department of Chemistry, The Pennsylvania State University, University Park, PA, 16802, USA

## Abstract

This work describes the development of a third-generation charge optimized many-body (COMB3) potential for Al-C and its application to the investigation of aluminum/graphene nanostructures. In particular, the new COMB3 potential was used to investigate the interactions of aluminum surfaces with pristine and defective graphene sheets. Classical molecular dynamics simulations were performed at temperatures of 300-900K to investigate the structural evolution of these interfaces. The results indicate that although the interfaces between Al and graphene are mostly weakly bonded, aluminum carbide can form under the right conditions, including the presence of vacancy defects in graphene, undercoordinated Al in surface regions with sharp boundaries, and at high temperatures. COMB3 potentials were further used to examine a new method to transfer graphene between Al surfaces as well as between Al and Cu surfaces by controlling the angle of the graphene between the two surfaces. The findings indicate that by controlling the peeling angles it is possible to transfer graphene without any damage from the surface having greater graphene/surface adhesion to another surface with less adhesion.

*Author to whom correspondence should be addressed. Email is sinnott@matse.psu.edu

## I. INTRODUCTION

Metal/graphene systems have been investigated for mechanical, electronic and energy applications in many configurations, including metal/graphene composites and nanostructures for electronic devices. [1-5] Among these, Al/graphene materials have attracted attention. For instance, the introduction of graphene to aluminum has been shown to overcome a number of limitations of monolithic Al and yield a desirable combination of strength, stiffness, toughness and density. [6] In addition, the optical, thermal and electrical properties of graphene mean that Al/graphene composites can produce new integrated structural-functional materials. [7] The interface between graphene and the metal surface plays a key role in determining the properties of these systems and their overall performance during applications such as catalysis, sensing and energy storage. Determining the structure-property relationships of these interfaces can further be beneficial to related systems such as carbon/metallic electrodes in nanoelectronics, the mechanism of graphene growth, and, more generally, organic-inorganic interface properties. [8] [9]

Several transmission electron microscopy (TEM) studies of graphene in aluminum matrix composites indicate that there is a sharp interface between Al and graphene without the formation of aluminum carbide. [10, 11] For example, Li *et al.* found that clean and strong interfaces are the result of metallurgical bonding on an atomic scale. [12] In contrast, other TEM studies report the formation of both stoichiometric and non-stochiometric aluminum carbide at Al/carbon interfaces. [13-15] Reduction in the strength of the graphene/Al-matrix composites was observed due to the formation of these aluminum carbide phases. [16, 17] There is additional experimental evidence that the improvement of mechanical properties such as tensile properties is due to the presence of limited amounts of thin carbide layers that increase graphene/Al interfacial bonding. [18] [19] Defect concentrations and temperatures reached in the experimental synthesis of graphene/Al composite may be related to the formation of aluminum carbide at the metal/graphene interface, but there is much about the structure-property relationships of graphene/Al interfaces that remains unknown.

Fundamental knowledge of graphene/metal substrate interfacial systems is additionally necessary to improve practical operations such as the transfer of intact graphene films cleanly from one surface to another following growth via chemical vapor deposition. [20][21] Since metal surfaces often serve as substrates on which graphene is grown, understanding the details of graphene/metal interfacial systems can help to develop improved methods for the clean and damage-free transfer of graphene between surfaces [22, 23]

Computational methods are highly complementary to experimental measurements and predict details of atomic-scale interactions at nanoscale interfaces. First principles, density functional theory (DFT) calculations predicted that the interface between Al (111) and graphene has a lattice mismatch that is less than 1.5%, and a corresponding binding energy is 0.027 eV per carbon atom. [24, 25] This is consistent with Al physisorbing to the graphene with van der Waals interactions. However, the sizes of the structures used in the DFT calculations were small, and external effects such as temperature and surface topology were not considered. Larger scale classical molecular dynamics (MD) simulations have not yet been widely employed due to the absence of reliable interatomic potentials for Al-C systems. Recently, a ReaxFF potential was developed to explore the effect of carbon coating on the oxidation of aluminum nanoparticles, [26] but the results for graphene/Al interfaces were not satisfactory since the Al-C interaction in this potential was not primarily fitted for such interfaces.

Here, we report the development of a reactive dynamic charge interatomic potential for Al-C interaction within the framework of the third generation of charge optimized many-body potential (COMB3) for Al and hydrocarbon [27][28][29][30] and apply it to the investigation of Al/graphene nanostructures. In particular, the potential is used in classical MD simulations to investigate the structural evolution of graphene/Al(111) interfaces and to consider the influence of the defects in the graphene sheet, the topology of the Al surface, and the temperature of the system. Additionally, we examine the way in which the angles between the graphene and Al at their interfaces affect the structural stability of the system; this leads to a new proposal on how to transfer graphene from one Al surface to another and from Cu to Al surfaces. The results of this work reveal changes in the bonding at Al/graphene interfaces with changing conditions and indicate the ways in which these interfacial interactions can be exploited to enable the reproducible manipulation of graphene sheets.

## II. COMPUTATIONAL DETAILS

## A. Development of Al-C COMB3 potential

The COMB3 potential is a variable charge, reactive interatomic empirical potential in which the total energy of the system is expressed as a sum of electrostatic energy ($U^{es}$), charge dependent short-range interactions ($U^{short}$), van der Waals interactions ($U^{vdW}$), and correction terms ($U^{corr}$) as shown in Eq. (1). Here $\{q\}$ and $\{r\}$ refer to the charge and coordinates of the atoms in the system, respectively. More details of the COMB3 potential and parameters can be found elsewhere. [27]

$$U^{total}[\{q\},\{r\}] = U^{es}[\{q\},\{r\}] + U^{short}[\{q\},\{r\}] + U^{vdW}[\{r\}] + U^{corr}[\{r\}] \qquad (1)$$

The parameterization of Al-C took advantages of the parameterization optimization software for materials (POSMat) [31] and an in-house code to minimize the cost function, which is a sum of weighted squared residuals between the calculated values and the training database. The training database was derived from previous and our own first-principle calculations.

Our first-principle calculations were performed using DFT as it is implemented in the Vienna *Ab initio* Simulation Package (VASP), using the generalized gradient approximation (GGA) and the Perdew-Burke-Ernzerhof (PBE) exchange-correlation functional. [32-34] The plane-wave cutoff energy was set to 520 eV and the Brillouin zones were sampled using a Monkhorst-Pack mesh with 4×4×4 k-points. The convergence criteria for geometry optimization were set at $1.0 \times 10^{-5}$ eV and $1.0 \times 10^{-3}$ eV·Å$^{-1}$ for energies and forces, respectively. The potential energy, interatomic forces, and the stress tensor were corrected to include van der Waals contributions using the vdw-DF2 method. [35]

## B. Classical molecular dynamics simulations

The classical MD simulations were carried out using the large-scale atomic/molecular massively parallel simulator (LAMMPS), an open source MD code made available by Sandia National Laboratory. [36] The geometric optimization was achieved using the conjugate gradient algorithm, and the sizes of structures were fully relaxed. The stopping tolerances for energies and forces were $1.0 \times 10^{-12}$ eV and $1.0 \times 10^{-10}$ eV·Å$^{-1}$, respectively, and time step was set at 0.1 fs. For dynamic simulations, a Nosé-Hoover barostat and a Langevin thermostat were applied to control the pressure and temperature of systems, respectively. Pressures within the simulations were set to 0 bar, and various temperatures of 300-900K were considered. The time step used in these simulations was 0.2 fs, and the simulations ran for 50 ps or until the energies of the systems fluctuated slightly about a constant value.

### III. POTENTIAL FITTING RESULTS

In the fitting process, the potential parameters were optimized for the interaction of atomic Al with graphene. Several structures were considered, including: one Al atom on the hexagonal, bridge and atop sites of pristine graphene; one Al atom or Al dimer on top of a single vacancy of graphene; two Al atoms on each side of a single vacancy of graphene; and one Al trimer on pristine graphene. The fitting results of the binding energy (BE) for each structure are illustrated in Table I. Most fitted COMB3 data is in general agreement with DFT results. The COMB3 BEs of one Al atom on different sites of pristine graphene are underestimated compared to DFT data. Since it is difficult to obtain perfect agreement between all the fitting data and database, such imperfection is retained to achieve overall good fitting results and to avoid overfitting.

Our potential parameters are also fitted to reproduce some properties of bulk crystalline structures. These fitting results are provided in Table II based on available published data and our own DFT calculations. This COMB3 potential accurately reproduced the heat of formation and lattice constant of bulk $Al_4C_3$. The heats of formation of Al-C in other structures in the database such as rutile, silica and NaCl and CsCl crystal structures are generally close to the DFT results, but structures such as fluorite and ZnS have lower heats of formation than predicted by DFT. Since the heats of formation of these structures are all above 0 eV/atom using either DFT or COMB3, these phases are predicted to be highly unstable and to thus decompose; therefore, this set of parameters is judged to be acceptable despite of the imperfect agreement with DFT data.

To further explore the ability of the potential to describe the energetics of phases not explicitly included in the fitting database, a genetic algorithm search for low-energy compounds were performed in the Al-C space. We used the genetic algorithm for structure prediction (GASP) package [37] coupled to LAMMPS for structure optimization and energy evaluation with COMB3. GASP enables grand-canonical searches for structures with different numbers of atoms in the unit cell and different compositions; the initial structures and compositions were chosen randomly. A more complete description of GASP is provided elsewhere.[37] The 0 K phase diagrams, also referred to as the convex hulls, are provided in Fig. 1. Bulk $Al_4C_3$ is correctly predicted to be the most stable Al-C phase with a formation energy of -0.09 eV/atom in agreement with DFT results. In addition to the $Al_4C_3$ phase, AlC and $Al_2C$ structures are predicted by COMB3 to have slightly negative heats of formation (-0.05 and -0.02 eV/atom); by contrast, DFT predicts positive heats of formation for these phases (0.08 and 0.11 eV/atom). Nonetheless, these two phases are still located above the convex hull indicating that COMB3 would still predict that they will both decompose to the lower energy $Al_4C_3$ structure. Given that the COMB3 potential has captured many properties for different Al-C phases, this imperfect parameter set is deemed to be acceptable despite these issues. This final set of newly developed parameters for Al-C interaction is seamlessly coupled with existing COMB3 potentials, including those for Al, Al-N, Al-O interactions. The full details of the parameterization are provided in Tables III and IV.

TABLE I. Binding energy (eV) of Al/graphene calculated by DFT and COMB3

| Structure* | DFT | COMB3 |
|---|---|---|
| 1 | -1.20 | -0.13 |
| 2 | -1.11 | -0.82 |
| 3 | -1.00 | -0.82 |
| 4 | -5.62 | -5.48 |
| 5 | -5.66 | -5.67 |
| 6 | -6.40 | -6.14 |
| 7 | -9.21 | -9.34 |
| 8 | -0.97 | -1.19 |

*Structure numbers:

(1) One Al at hexagonal site of pristine graphene

(2) One Al at bridge site of pristine graphene

(3) One Al at atop site of pristine graphene

(4) One Al at single vacancy of graphene

(5) One Al dimer vertically at single vacancy of graphene

(6) One Al dimer horizontally at single vacancy of graphene

(7) Two Al atoms on each side of single vacancy of graphene

(8) One Al trimer horizontally on pristine graphene.

TABLE II. Properties of crystalline phases from DFT and COMB3 calculations

| Property | DFT | COMB3 |
|---|---|---|
| $Al_4C_3$ bulk phase | | |
| $\Delta H_f$ (eV/atom) | -0.09 | -0.09 |
| a (Å) | 3.36 | 3.39 |
| c (Å) | 25.12 | 24.66 |
| $\Delta H_f$ (eV/atom) of other phases | | |
| [NaCl] | 0.97 | 1.11 |
| [ZnS] | 1.25 | 0.63 |
| [CsCl] | 2.63 | 2.35 |
| silica | 1.47 | 1.41 |
| rutile | 1.00 | 0.86 |
| fluorite | 2.66 | 0.54 |

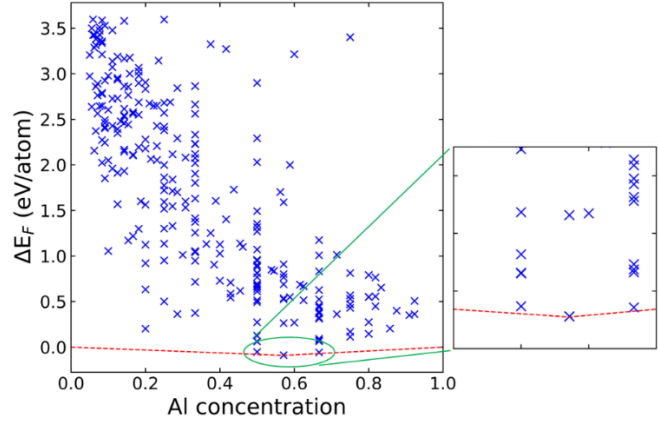

FIG. 1. Al-C phase diagram at 0 K generated by a genetic algorithm search with COMB3 potential. Each data point represents a unique bulk phase.

TABLE III. Two-body parameters for the Al-C COMB3 potential

| Parameter | C-Al | Al-C |
|---|---|---|
| $A_{ij}$ | 838.2192 | 838.2192 |
| $B^1_{ij}$ | 118.7144 | 118.7144 |
| $B^2_{ij}$ | 0.0 | 0.0 |
| $\lambda$ | 2.9071 | 2.9071 |
| $\alpha^1_{ij}$ | 1.5377 | 1.5377 |
| $\alpha^2_{ij}$ | 0.0 | 0.0 |
| $\beta$ | 0.0 | 3.8069 |
| $b_6$ | 0.0 | 0.0 |
| $b_5$ | 0.0 | 0.0 |
| $b_4$ | 0.4206 | 1.8854 |
| $b_3$ | 0.5092 | 0.9078 |
| $b_2$ | 0.0025 | 0.2242 |
| $b_1$ | -0.0692 | 0.0548 |
| $b_0$ | 0.0650 | 0.0075 |
| $R^{min}$ | 2.9000 | 2.9000 |
| $R^{max}$ | 3.2000 | 3.2000 |
| cosm1 | 0.9929 | 0.3269 |
| cosm2 | 2.9131 | 0.8942 |

TABLE IV. Three-body parameters for the Al-C COMB3 potential

| Parameter | Al-C-Al | C-Al-C | C-Al-Al | C-C-Al | Al-C-C |
|---|---|---|---|---|---|
| LP$_0$ | 0.2584 | -0.0786 | -0.0106 | 0.6898 | -0.8963 |
| LP$_1$ | 0.0 | 0.0 | 0.0 | 1.3160 | -2.4944 |
| LP$_2$ | 0.0 | 0.0 | 0.0 | 2.4836 | -1.1926 |
| LP$_3$ | 0.0 | 0.0 | 0.0 | 1.0533 | 0.0 |
| LP$_4$ | 0.0 | 0.0 | 0.0 | 0.6482 | 0.0 |
| LP$_5$ | 0.0 | 0.0 | 0.0 | 0.0 | 0.0 |
| LP$_6$ | 0.0 | 0.0 | 0.0 | 0.0 | 0.0 |

## IV. GRAPHENE/ALUMINUM SYSTEMS

### A. Work of adhesion for Al/graphene

An interface between an Al (111) surface and pristine graphene was constructed as discussed in previous studies [24, 25]; a snapshot is provided in Fig. 2. We first explored the dependence of the energy of this system on the number of Al layers. The work of adhesion ($W_a$) is defined in Eq. (2) where $E_{int}$ is the energy of relaxed Al/graphene interface structure, $A$ is the area of the interface, $E_G$ and $E_{Al}$ are the energies of graphene and aluminum slab after they are isolated, respectively.

$$W_a = (E_G + E_{Al} - E_{int})/A \qquad (2)$$

The work of adhesion of graphene on various numbers of Al layers as predicted by COMB3 is illustrated in Fig. 3. $W_a$ increases slightly as the number of Al layers increase, which is similar to behavior predicted for other similar systems. [38] $W_a$ converges to around 0.093 eV per C atom (0.55 J/m$^2$) in the interface. The $W_a$ between graphene and six layers of Al here is 0.076 eV/C atom, which is in the same range predicted by previous DFT calculations of 0.027 eV/C atom in the interface. [25] The overestimation of the COMB3 result is due to the slightly overestimation Columbic contribution to the total energy. The predicted interfacial spacing between Al and graphene is around 0.34 nm which is in good agreement with DFT. [25]

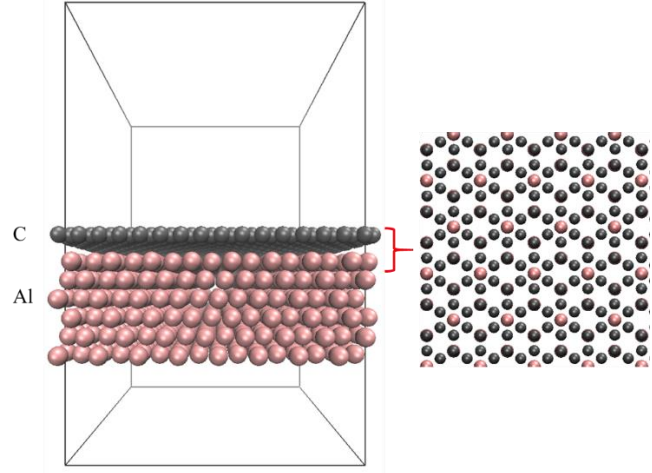

FIG. 2. Side and top views of the interface between Al (111) and pristine graphene (0001)

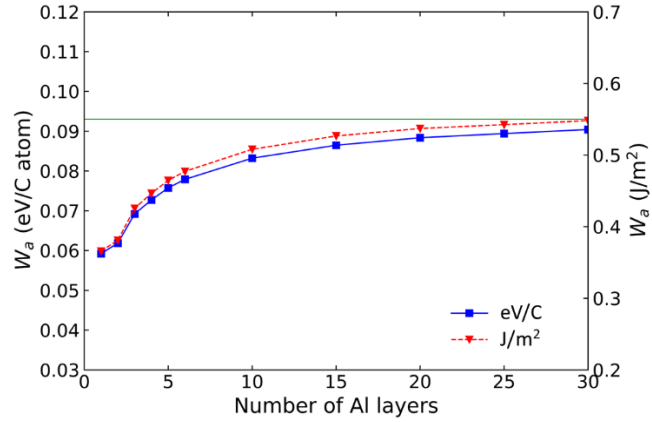

FIG. 3. Work of adhesion for the Al/graphene interface illustrated in Fig. 2 as a function of the number of Al layers.

### B. Carbide formation at the Al/graphene interface

Next, we considered the influence of defects in graphene, the surface topology of Al, and system temperature on the possible formation of aluminum carbide at Al/graphene interfaces. Three types of defects in graphene were considered: single vacancy (SV), di-vacancy (DV) and di-interstitial (DI), as illustrated in Fig. 4. The results suggest that the SV has little impact on the $W_a$ of graphene/Al interface. The $W_a$ is slightly increased by DV and decreased by DI on graphene, but the order of magnitude of their influence is small as shown in Table V.

Classical MD simulations were performed with COMB3 on these interfaces at 300, 600 and 900K to investigate their structural stability. For both pristine and defective graphene, their structures were stable and their interfaces with Al remained flat and well-defined at all these temperatures. The interaction between graphene and Al remained physisorption, and there was no new Al-C bond observed at the graphene/Al interfaces in all cases. This suggests that simple defects in graphene on flat Al surfaces under various external temperatures are not significant enough to result in chemical reactions between C and Al.

The influence of the metal surface topology, such as steps and islands, was considered and the structural evolution of the graphene/Al interfaces considered at temperatures of 300-900K. The results are provided in Fig. 5. The simulations predict that no new Al-C bonds are formed at any of the interfaces between Al and defect-free graphene for any temperatures considered. The reaction between Al and pristine graphene is so inert that the topology of Al surface and external temperature do not lead to carbide formation. However, the formation of Al-C bonds is predicted between the DI defect of graphene and the tip Al atoms of the island on Al surface at 900K. For SV and DV defects, we predicted that new Al-C bonds are also formed at all the temperatures considered here. These results suggest that the formation of aluminum carbide is most likely initiated by new Al-C bonds between the C in the defect of graphene and the Al of those regions with sharp boundary on the Al surface. Other effects such as mechanical forces and environmental factors are not considered in this work.

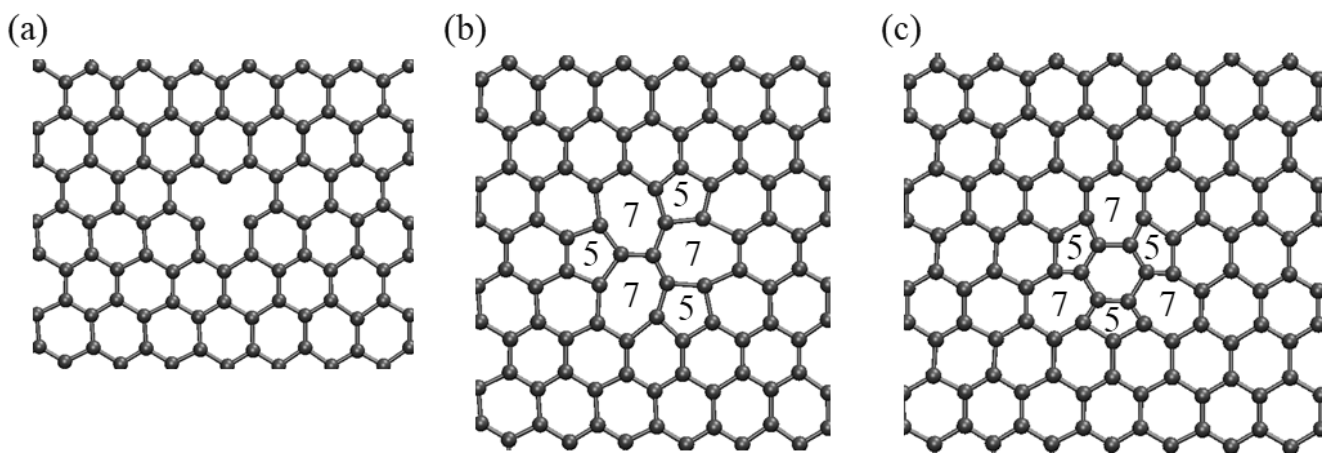

FIG. 4. Defects in graphene (top view): (a) single vacancy, (b) di-vacancy and (c) di-interstitial. The numbers indicate the sizes of the carbon rings.

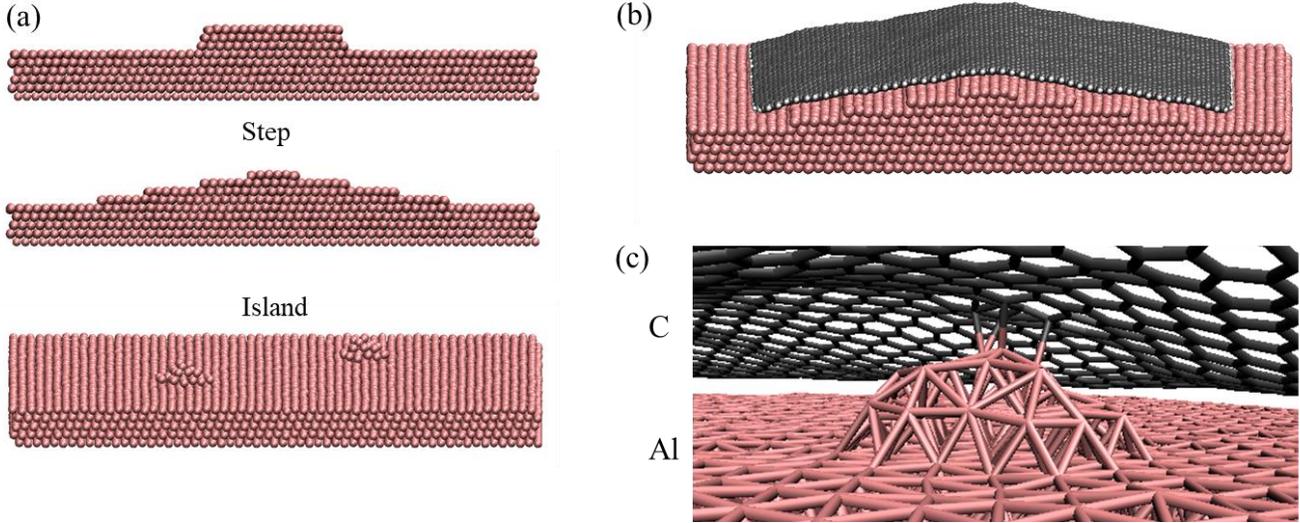

FIG. 5. Snapshots of (a) Al surface substrates with different surface topology, (b) optimized structures of graphene/Al with steps on the Al(111) surface, and (c) the newly formed Al-C bonds at the interface.

TABLE V. Work of adhesion (eV/C atom) for Al layers on different types of graphene structures

| Al layers | Graphene types | | | |
| --- | --- | --- | --- | --- |
| | Pristine | SV | DV | DI |
| 6 | 0.076 | 0.075 | 0.079 | 0.071 |
| 20 | 0.089 | 0.090 | 0.095 | 0.086 |

## C. Graphene transfer

Kendall and co-workers [39] predicted that when a thin film of width $w$ is peeled off from a substrate, smaller peeling forces, $F$, along the thin-film plane are needed for larger peeling angles, $\theta$, between the thin film and substrate. Kendall's peeling off model is given by the equation below:

$$Fw^{-1}(1 - \cos \theta) - R = 0 \qquad (3)$$

where $R$ is the adhesion energy between the film and substrate. Kuznetsov et al., for example, successfully explored this angle dependence of the peeling forces to propose a model for accumulation of connections between bundles of carbon nanotubes (CNTs) in a CNT forest during a pulling out process of formation of CNT macroscopic fibers. [40] On the basis of Kendall's model, when a thin film has its ends laid on two surfaces and a tension force is applied to the film, the film will detach from the surfaces with smaller peeling forces.

In this work, a similar idea is considered in a new proposal for a method to transfer graphene from one surface to another. We focus on the transfer of graphene from a surface having larger graphene/surface adhesion to another having smaller adhesion. We first studied a system in which two Al surfaces are connected by a curved graphene sheet. The system was equilibrated using the COMB3 potential in MD simulations at 300K to optimize the configuration of the graphene, as indicated in Fig. 6. Two identical Al(111) surfaces were placed in parallel in the simulation box, and a graphene sheet was in contact with both of them such that the initial angle between graphene and Al at the intersection was 60°. The graphene carbon atoms closest to the Al are bonded by physisorption, and its curvature minimizes the local stress within the graphene sheet.

Next, the top surface was rotated by 30° clockwise, in Fig. 7(a), and counterclockwise, in Fig. 7(b), respectively, and was pulled upward at a rate of 0.5 Å/ps. When the peeling angle of the bottom substrate ($\theta_1$) is smaller than that of the top substrate ($\theta_2$), larger forces are required to separate the graphene sheet from the bottom graphene/Al interface. As a result, the graphene sheet is gradually peeled off from the top graphene/Al interface, as illustrated in Fig. 7(a). In contrast, when the peeling angle with the top surface is smaller, the graphene sheet remains attached with the top surface and is gradually peeled off from the bottom surface. This suggests that larger peeling angles reduce the structural stability of graphene/metal interfaces, which should be noted in the design of graphene/metal systems. These results provide guidance for the transfer of graphene sheets from one substrate to another. Although there are mechanical models that examine the structure and stability of thin films during the complete peeling off process [41] the difference in peeling off angles when the film is laid on two surfaces has yet to be considered.

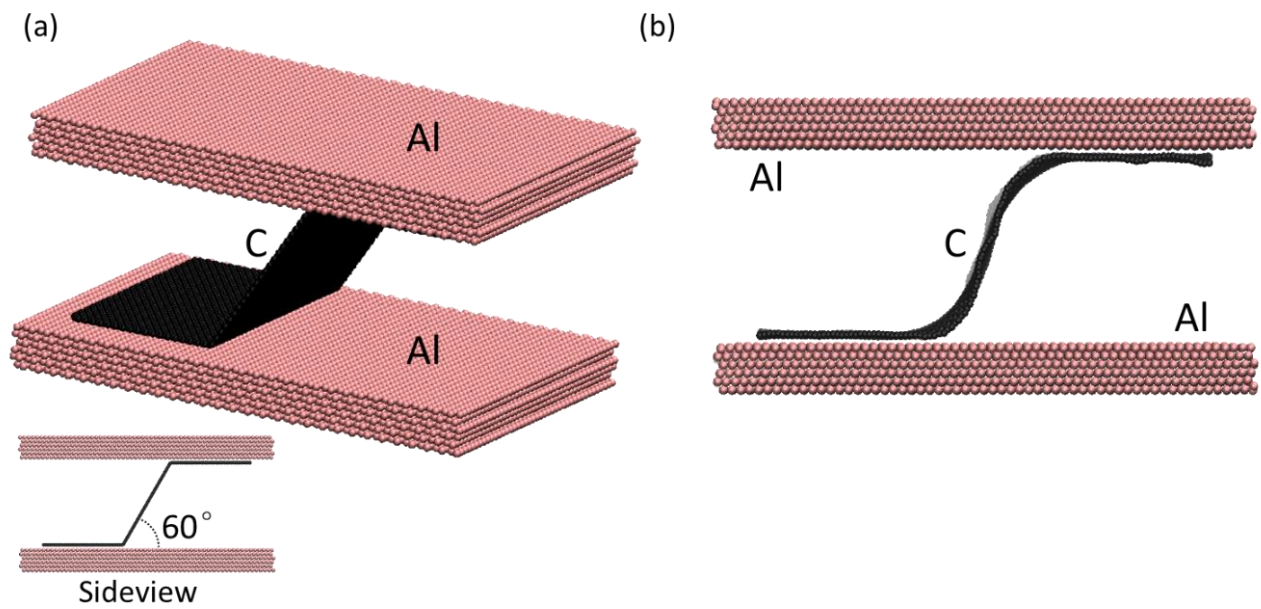

FIG. 6. An example of the Al/graphene/Al systems: (a) initial structures as-built and (b) the COMB3-optimized structures.

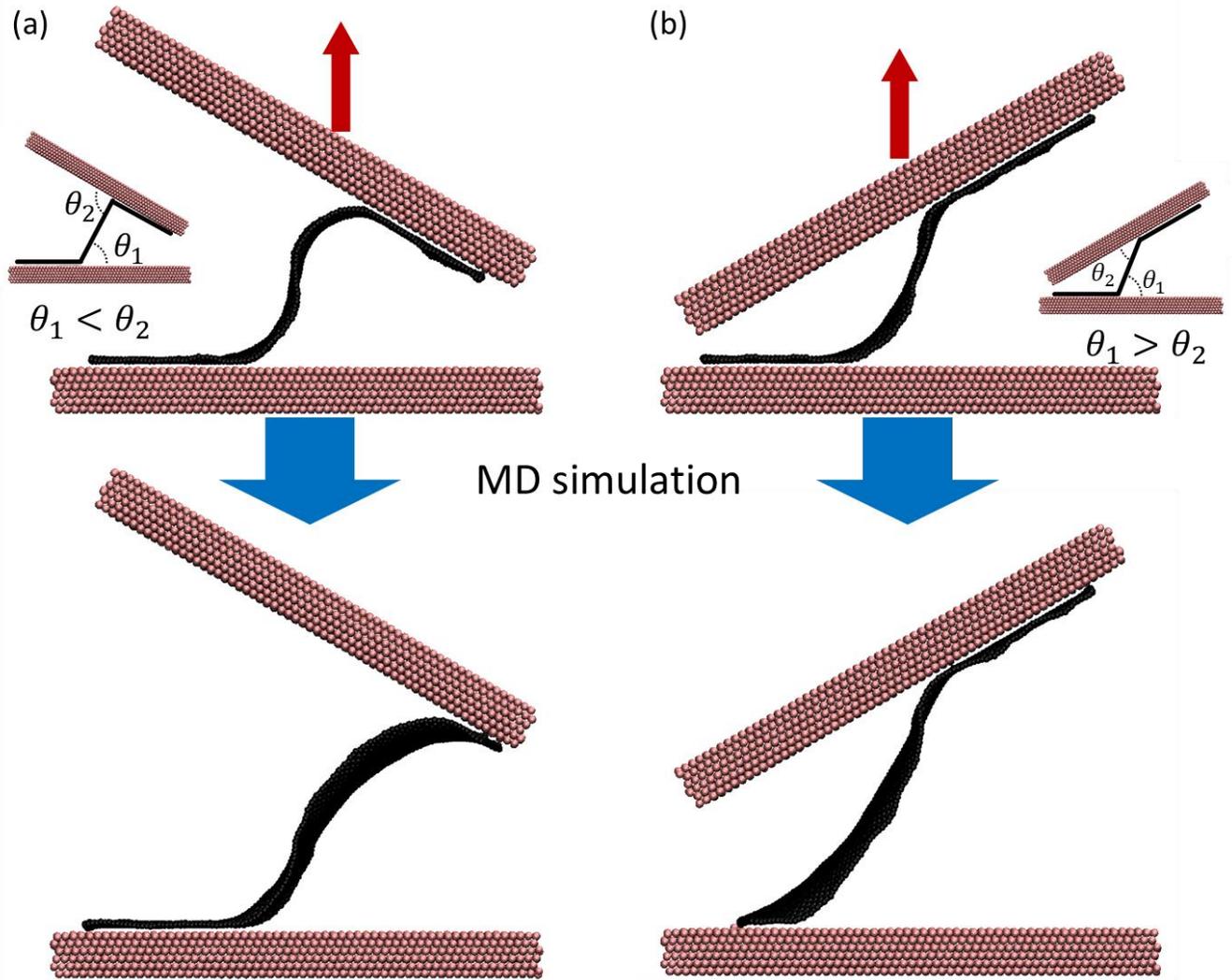

FIG. 7. Snapshots for systems with different peeling angles: (a) top angle larger than bottom angle and (b) top angle smaller than bottom angle. The red arrows indicate the peeling direction of top surface.

To further explore the application on transferring graphene between surfaces, we investigated a similar process using two different surfaces connected by the curved graphene sheet, one of aluminum and one of copper. The results are provided in Fig. 8; the details of the COMB3 potential for Cu-C interactions may be found elsewhere [30]. Based on Kendall's model, the condition of peeling off angles in favor of transferring graphene from a surface having larger adhesion to one have smaller adhesion can be estimated. Assuming $R_{LARGE}$ and $\theta_{LARGE}$ ($R_{LOW}$ and $\theta_{LOW}$) as the adhesion energies and the peeling off angles between graphene and the surface with larger (low) adhesions, respectively, we can obtain equations:

$$F_{LOW}\, w^{-1}\, (1 - \cos \theta_{LOW}) - R_{LOW} = 0 \tag{4a}$$

$$F_{LARGE}\, w^{-1}\, (1 - \cos \theta_{LARGE}) - R_{LARGE} = 0 \tag{4b}$$

where $F_{LARGE}$ and $F_{LOW}$ are the forces of tension on the graphene laid on each surface, respectively, during a peeling off process. In the process illustrated in Fig. 8, we define r as the ratio $R_{LOW}/R_{LARGE}$ so that $r < 1$. To transfer graphene from the surface with larger adhesion to one with smaller adhesion, the force needed to peel graphene off the surface with larger adhesion must be smaller than that needed for the other surface. Therefore, when the relation $F_{LOW} > F_{LARGE}$ is satisfied, we can obtain equation (5) by using $R_{LOW} = r\, R_{LARGE}$.

$$\cos \theta_{LOW} > 1 - r + r \cos \theta_{LARGE} \tag{5}$$

When the graphene sheet has the same peeling angle with both the Al(111) and the Cu(111) surfaces, eq. (5) is not satisfied and the graphene sheet remains attached to the Cu because of its stronger work of adhesion of 0.104 versus 0.076 eV/C with the Al. It is not surprising that the graphene is eventually peeled from the Al surface. The right side of eq. (5) shows a critical value for the peeling angle of the surface having lower adhesion. It may not be possible to transfer graphene from the Cu surface (higher adhesion) to Al surface (lower adhesion) when such condition is not met. In contrast, if the peeling angle of graphene/Al is smaller enough than that of graphene/Cu, the graphene can be gradually peeled from the Cu surface and remains attached to the Al surface. When the peeling off angle for graphene/Cu is 60° ($\theta_{LARGE}$), the critical peeling angle for graphene/Al is ~ 50.6° for $r = 0.73$ in this work. Here, we have chosen $\theta_{LOW} = 30°$ for graphene/Al to evaluate our prediction if the critical condition is satisfied. This result indicates that by controlling the peeling angles of graphene with the surfaces, it is possible to transfer graphene from a surface with stronger interaction to a surface with a weaker interaction.

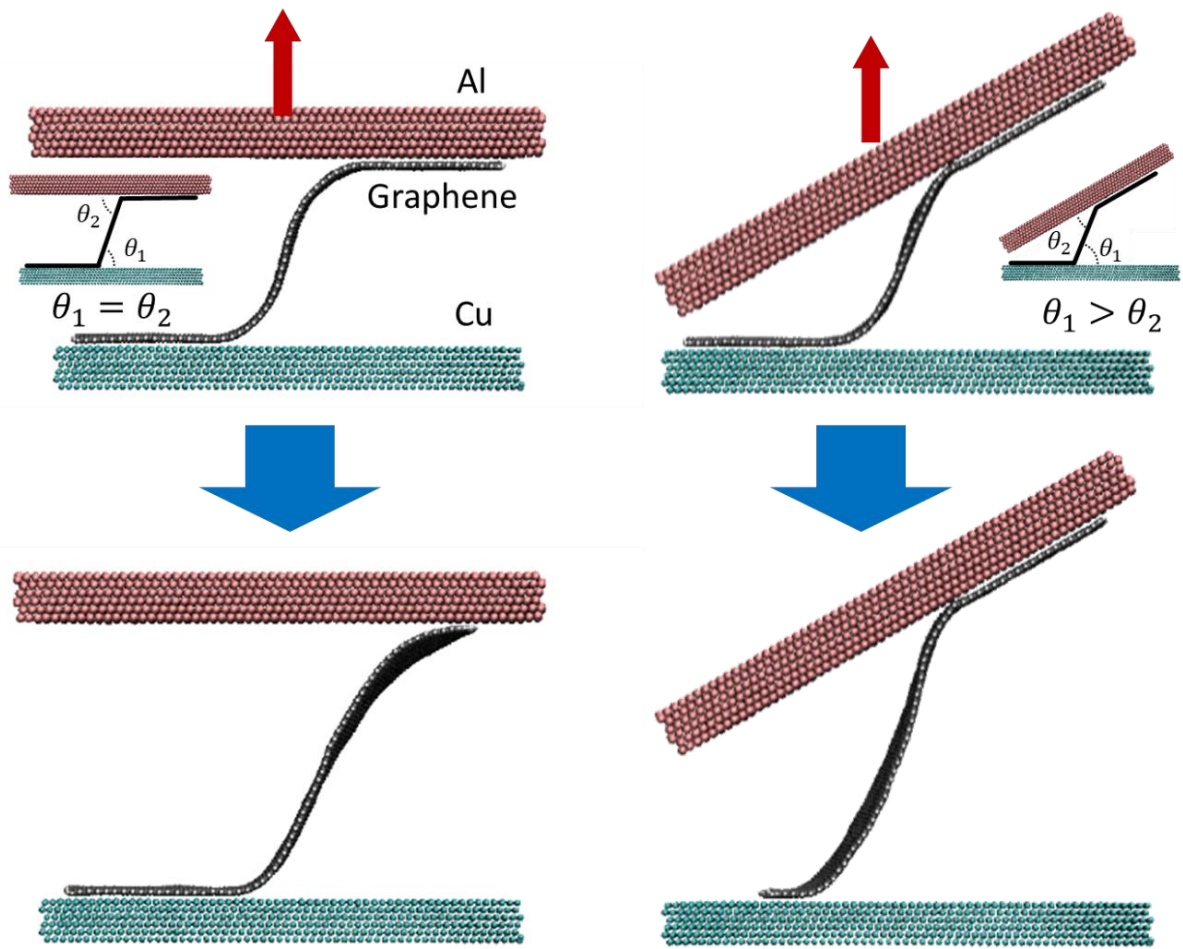

FIG. 8. Snapshots of the graphene sheet connected with Al(111) and Cu(111) surfaces. The red arrows indicate the peeling direction of the top surface in the MD simulations.

**CONCLUSIONS**

In this work, we developed a third-generation charge optimized many-body potential for Al-C interactions for use in classical MD simulations of Al/graphene nanostructures. This potential can reproduce results in agreement with most DFT data of the binding energy of Al with graphene and the heat of formation for crystalline bulk phase. It can also appropriately generate convex hull for Al-C phase space.

Using this COMB3 potential, we investigated various graphene/Al interfaces at different temperatures in classical molecular dynamics simulations. No new Al-C bond formation has been observed at the graphene/Al interfaces for both pristine and defective graphene under all temperatures when the Al surface is flat, but we found new Al-C bond formed in the Al/graphene interfaces between the C in defects of graphene and undercoordinated Al on the sharp boundaries of Al surfaces. Results suggest that the formation of aluminum carbide can be aided by the vacancy defects of graphene, the undercoordinated Al on sharp ends of Al surfaces and high external temperatures in the experimental samples. In addition, since larger peeling forces are needed to remove a graphene from the graphene/Al interfaces with smaller peeling angles, a graphene sheet can be separated from one graphene/Al interface with larger peeling angle to another interface with smaller peeling angles. Taking advantages of such difference between peeling angle, graphene can be transferred from Cu to Al when the peeling angle of graphene/Al interface is smaller enough than graphene/Cu interface due to the larger force required to separate the graphene/Al interfaces.

This work provides fundamental structure-property details to different graphene/Al interfaces occurred in experiments and illustrates a method that can be used to transfer graphene between different substrates. We hope our work can aid better understanding of graphene/Al interfaces in the experimental design and the development of graphene-based materials.

## ACKNOWLEDGEMENTS

This work is supported by UNCAGE-ME, an Energy Frontier Research Center funded by the U.S. Department of Energy, Office of Science, Basic Energy Sciences under Award #DE-SC0012577. All computations and simulations are performed on the ICS-ACI supercomputers at the Pennsylvania State University. A.F.F. acknowledges support from the Brazilian Agencies CNPq (grant number #311587/2018-6) and São Paulo Research Foundation (FAPESP, grant #2018/02992-4), and from FAEPEX/UNICAMP.